\documentclass[12pt,a4paper]{elsart}
\usepackage{graphicx}
\usepackage{latexsym}
\usepackage{amssymb}
\usepackage{ulem}

\def\Fig{Figure}

\def\Tab{Table}

\def\echemAV{\ensuremath{\overline{E}_{\rm chem}}}

\def\note#1{{\it Note: #1}}
\begin{document}
\begin{frontmatter}

  
  \title{On-surface and Subsurface Adsorption of Oxygen on Stepped
    Ag(210) and Ag(410) Surfaces}

\author{A. Kokalj$^{a,b}$},
\author{N. Bonini$^{a}$\corauthref{cor}},
\ead{bonini@sissa.it}
\corauth[cor]{Corresponding Author. Phone: +39-040 3787477,
Fax: +39-040-3787528.}
\author{A. Dal Corso$^{a}$},
\author{S. de Gironcoli$^{a}$}\and
\author{S. Baroni$^{a}$}

\address{$^a$SISSA--Scuola Internazionale Superiore di Studi
  Avanzati and INFM {\it DEMOCRITOS} National Simulation Center,
  I-34014 Trieste, Italy}
\address{$^b$Jo\v zef Stefan Institute, SI-1000 Ljubljana, Slovenia}

\begin{abstract}
  The adsorption of atomic oxygen and its inclusion into subsurface
  sites on Ag(210) and Ag(410) surfaces have been investigated using
  density functional theory.
  We find that---in the absence of adatoms on the first metal
  layer---subsurface adsorption results in strong lattice distortion
  which makes it energetically unfavoured. 
  However subsurface sites are significantly stabilised when a
  sufficient amount of O adatoms is present on the surface.
  At high enough O coverage on the Ag(210) surface the mixed
  on-surface + subsurface O adsorption is energetically favoured
  with respect to the on-surface only adsorption.
  Instead, on the Ag(410) surface, at the coverage we have considered
  (3/8 ML), the existence of stable terrace sites makes the subsurface
  O incorporation less favourable. These findings are compatible with
  the results of recent HREEL experiments which have actually motivated
  this work.

\end{abstract}

\end{frontmatter}

\linespread{1.0}
%
%
\section{Introduction}
\label{sec:Intro}

The interaction between oxygen and silver surfaces has been the
subject of extensive experimental and theoretical research because of
its key role in important heterogeneous catalytic
reactions, such as partial oxidation of methanol to formaldehyde or
ethylene epoxidation \cite{vanSanten_AC35}.  Despite significant efforts,
these systems are still not well understood and the identification of the
active species involved in these reactions remains unclear.  In
particular, defect sites and subsurface oxygen species have attracted
considerable attention since they are believed to be important in the
ethylene epoxidation reaction \cite{vandenhoek_JPC93}.

In recent experimental works
\cite{savio_PRL87,savio_JPCM14,vattuone_PRL90} Rocca and coworkers
have investigated the interaction of oxygen with Ag(410) and Ag(210)
surfaces with the aim of understanding the role of steps in the
adsorption of oxygen on silver.  In particular, using a supersonic
molecular beam to dose oxygen on the surface and characterising the
final adsorption state by vibrational spectroscopy (HREEL), they have
studied the different oxygen species which form on these surfaces.  Of
particular interest are the results for O/Ag(210)
\cite{vattuone_PRL90}.  The HREEL spectra of this system show a peak
around 56 meV which is proposed to be due to the vibration of an
oxygen atom occupying a subsurface site. Such a peak appears only when
preadsorbed oxygen is present on the surface and it is not observed
neither on low Miller index surfaces nor on O/Ag(410).

Motivated by these results we have performed an {\it ab initio}
calculation to investigate the possibility of incorporating oxygen
into interstitial subsurface sites on Ag(210) and Ag(410)
surfaces.

Recent theoretical works have studied the inclusion of O into
subsurface regions of low Miller index silver surfaces, i.e.  Ag(111)
\cite{Li_PRB67} and Ag(100) \cite{O-Ag001}.  It was found that pure
subsurface O atoms are always unstable with respect to the oxygen
molecule but the presence of co-adsorbed on-surface O atoms can
stabilise them.  More generally, we find that the adsorption in
subsurface sites depends significantly on coverage. Isolated
subsurface oxygen atoms are always less stable than adatoms occupying
favourable on-surface sites, essentially because of the high energy
cost for lattice distortion induced by subsurface O incorporation.
However, increasing the coverage the preference of on-surface sites
decreases and, for example, in O/Ag(111) pure subsurface sites are
stabilised at coverages higher than 0.5 ML \cite{Li_PRB67}.  This
behaviour is due to the fact that the repulsive interaction between
the adsorbates is more effectively shielded for the subsurface atoms
than for the on-surface ones.  In this work we consider high Miller
index surfaces in order to investigate the effects of steps on the
incorporation of oxygen atoms into subsurface sites.

%
%
\section{Computational Details}
\label{sec:Method}

Calculations were performed in the framework of density functional
theory (DFT) using the generalised gradient approximation (GGA) of
Perdew-Burke-Ernzerhof (PBE)~\cite{Perdew_PRL77}. We have used the
pseudopotential method with ultra-soft pseudopotentials
\cite{Vanderbilt_PRB41} and plane-wave basis sets up to a
kinetic-energy cutoff of 27 Ry (216 Ry for the charge-density).
Details about the Ag and O pseudopotentials are reported in
Ref.~\cite{Cipriani_SS501}. Brillouin zone (BZ) integration has been
performed with the Gaussian-smearing special-point technique
\cite{MonkhorstPack_PRB13,Methfessel_PRB40} with a smearing parameter
of 0.03 Ry. Calculations have been done using the
{\tt PWscf} package \cite{PWSCF_WEB}, while the molecular graphics
were produced by the {\tt XCRYSDEN} \cite{Kokalj_JMGM17}
graphical package.

We have used periodic super-cells as models of the surface. The
Ag(210) surface is modelled with slab of 14 (210) layers, which
corresponds to seven (100) layers rotated by an angle $\phi_2$
($\tan(\phi_2)=1/2$) around the [001] axis and merged together so as
to form the Ag(210) surface. The Ag(410) is modelled with slabs of 20
(410) layers corresponding to five (100) layers rotated by an angle
$\phi_4$ ($\tan(\phi_4)=1/4$) and merged together.  Adjacent slabs are
separated by a vacuum region of at least 16 a.u.  Oxygen is adsorbed
on both sides of the slab and all the structures have been fully
relaxed until the Hellmann-Feynman forces were lower than $10^{-3}$
Ry/a.u.

Adsorption of atomic oxygen has been modelled by (2$\times$1) surface
super-cells. For the Ag(210) surface the BZ integrations have been
performed using a (4$\times$7$\times$1) Monkhorst-Pack mesh, while for
the Ag(410) surface a ($4\times4\times1$) mesh has been applied. All the
calculations are non-spin-polarised, because the interaction between O
and the Ag substrate results in a negligible spin moment on the O atom.

The chemisorption energies, \echemAV, are referred to the clean
Ag($n$10) surface ($n=2,4$) and the isolated oxygen molecule,
$\echemAV = (E_{\rm O/Ag} - E_{\rm Ag} - N_{\rm O}(E_{\rm
O_2}/2))/N_{\rm O}$, where the total energy of the
adsorbate--substrate system, of the clean surface, of the isolated
O$_2$ molecule, and the number of adsorbed oxygen atoms are
represented by $E_{\rm O/Ag}$, $E_{\rm Ag}$, $E_{\rm O_2}$ and $N_{\rm
O}$, respectively.  $E_{\rm O_2}$ is estimated from a spin-polarised
calculation. With this definition, stable adsorbates have negative
chemisorption energies.

%
%
\section{Results and Discussion}
\label{sec:Results}

\begin{figure}[htbp]
  \centering
  \linespread{1.0}
  \begin{tabular}[t]{cc}
    \footnotesize ~~~~Ag(210) &
    \footnotesize ~~~~Ag(410)\\
    \includegraphics*[height=0.25\textwidth]{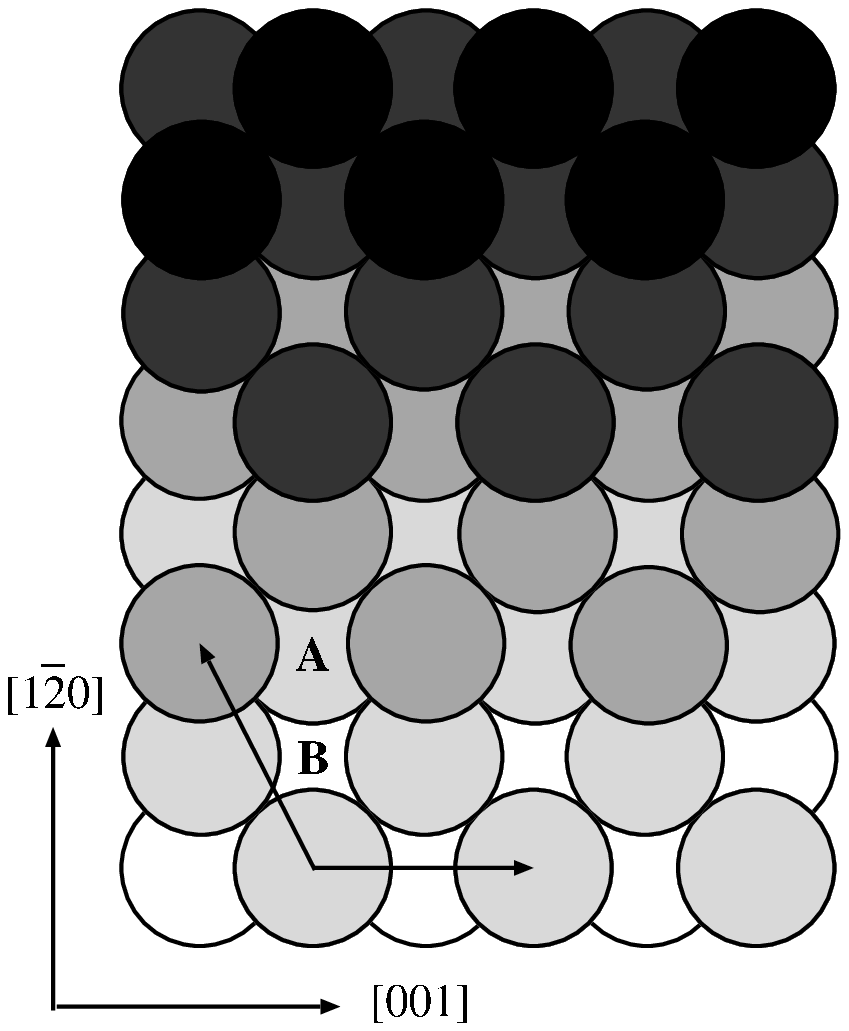}&
    \includegraphics*[height=0.25\textwidth]{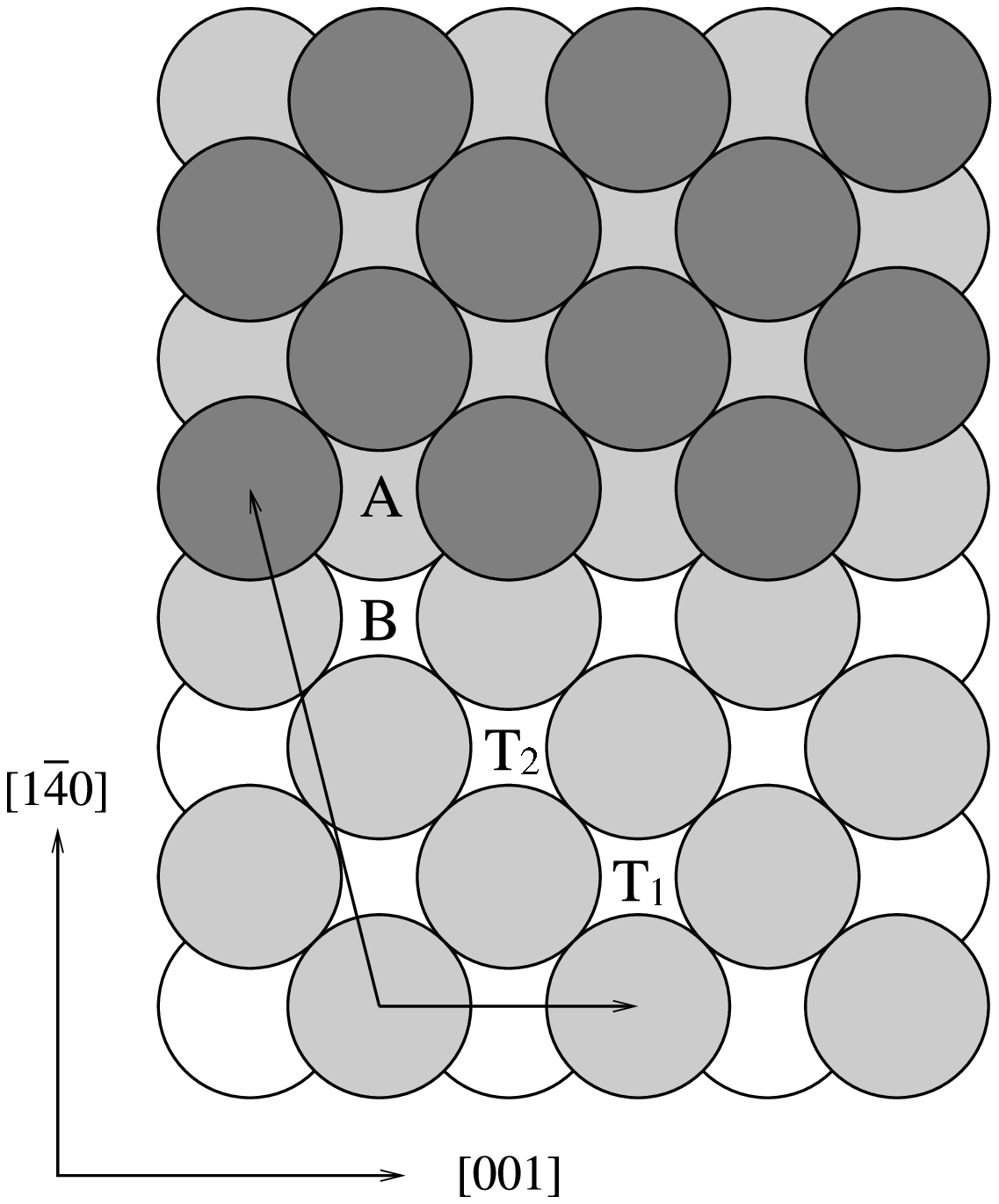}\\
  \end{tabular}
  \caption{Schematic representation of the Ag(210) (left) and Ag(410)
    surface (right). The labels A, B, T$_1$ and T$_2$ indicate
    different adsorption sites: A, above the step; B, below the step;
    T$_1$ and T$_2$, hollow sites on terrace.  The surface
    unit-cell vectors, and the [$001$] and [$1\bar{n}0$], $n=4,2$,
    crystal axes are also shown.}
\label{fig:surf}
\end{figure}

In previous works on Ag(001) \cite{Cipriani_SS501,Loffreda_SS530}, we
found that the surface four-fold hollow site is the most stable one
and the chemisorption energy was found to decrease increasing the
coverage, a fact related to the repulsive lateral electrostatic
interaction between negatively charged oxygen adatoms.  Recently, we
performed a systematic study of on-surface atomic oxygen adsorption on
Ag(410) and Ag(210) surfaces \cite{Bonini_subm}.  \Fig~\ref{fig:surf}
shows the structure of the two surfaces and the on-surface sites that
have been studied.  In site \textsc{A} the oxygen adatom lies just
above the step and is coordinated with three Ag surface atoms; in site
\textsc{B} the oxygen adatom lies in the hollow site just below the
step. These two sites are present on both surfaces.  The Ag(410)
surface possesses also the \textsc{T$_1$} and \textsc{T$_2$} sites
(see \Fig~\ref{fig:surf}) with the O adatom in the four-fold hollow
sites on the (100) terrace.  We find that the configuration in which
the adatoms occupy the sites A forming --O--Ag--O--Ag-- rows
(configuration labelled A--A) is particularly stable.  On the Ag(210)
surface the step decoration is significantly more favourable than the
other adsorption configurations.  On Ag(410), instead, the terrace
sites have a chemisorption energy almost degenerate with that of the
step sites. In particular, on this surface, when the adatoms are far
apart from each other the hollow sites on terrace are slightly more
stable than the step edge sites, while at higher coverage the adatoms
slightly prefer to decorate the steps.

Let us focus now on the energetics of oxygen inclusion into subsurface
interstitial sites, addressing in particular the effects of on-surface
co-adsorption.  Our results are summarised in \Tab~\ref{tab:echem} and
discussed in the next paragraphs.

%
%
\begin{table}
  \begin{center}
    \caption{Average chemisorption energies, \echemAV, for
      purely subsurface, mixed on-surface + subsurface, and on-surface
      only adsorption of atomic oxygen on Ag(210) and Ag(410)
      surfaces. The label of \textsc{Octa}, \textsc{Tetra},
      \textsc{A}, \textsc{B}, and \textsc{T$_1$} are defined in text.}

    \label{tab:echem}
    \linespread{1.0}
    \def\arraystretch{1.0}
\begin{tabular}{l l l c}
\hline
Surface   & Coverage        & Configuration &  \echemAV (eV) \\
\hline
Ag(210)   &\\
          & $\theta=1/4$~ML & \textsc{Octa}        & $-$0.07\\
          &                 & \textsc{Tetra}       & $-$0.14\\
          &                 & \textsc{A}           & $-$0.68\\[1em]

          & $\theta=3/4$~ML & \textsc{A--A + Octa} & $-$0.54\\
          &                 & \textsc{A--A + B}    & $-$0.40\\[1em]
\hline
Ag(410)   &\\
          & $\theta=3/8$~ML & \textsc{A--A + Octa} & $-$0.55\\
          &                 & \textsc{A--A + B}    & $-$0.68\\
          &                 & \textsc{A--A + T$_1$}   & $-$0.63\\[1em]
          \hline
\end{tabular}
\end{center}
\end{table}

\begin{figure}[htbp]
  \centering
  \linespread{1.0}
  \def\arraystretch{1.0}
  \begin{tabular}[t]{cc}
    \multicolumn{2}{c}{\it O atom in octahedral-subsurface site}\\[-0.5em]
    \scriptsize\textsc{Initial structure} &
    \scriptsize\textsc{Optimised structure}\\
    \includegraphics[height=0.25\textwidth]{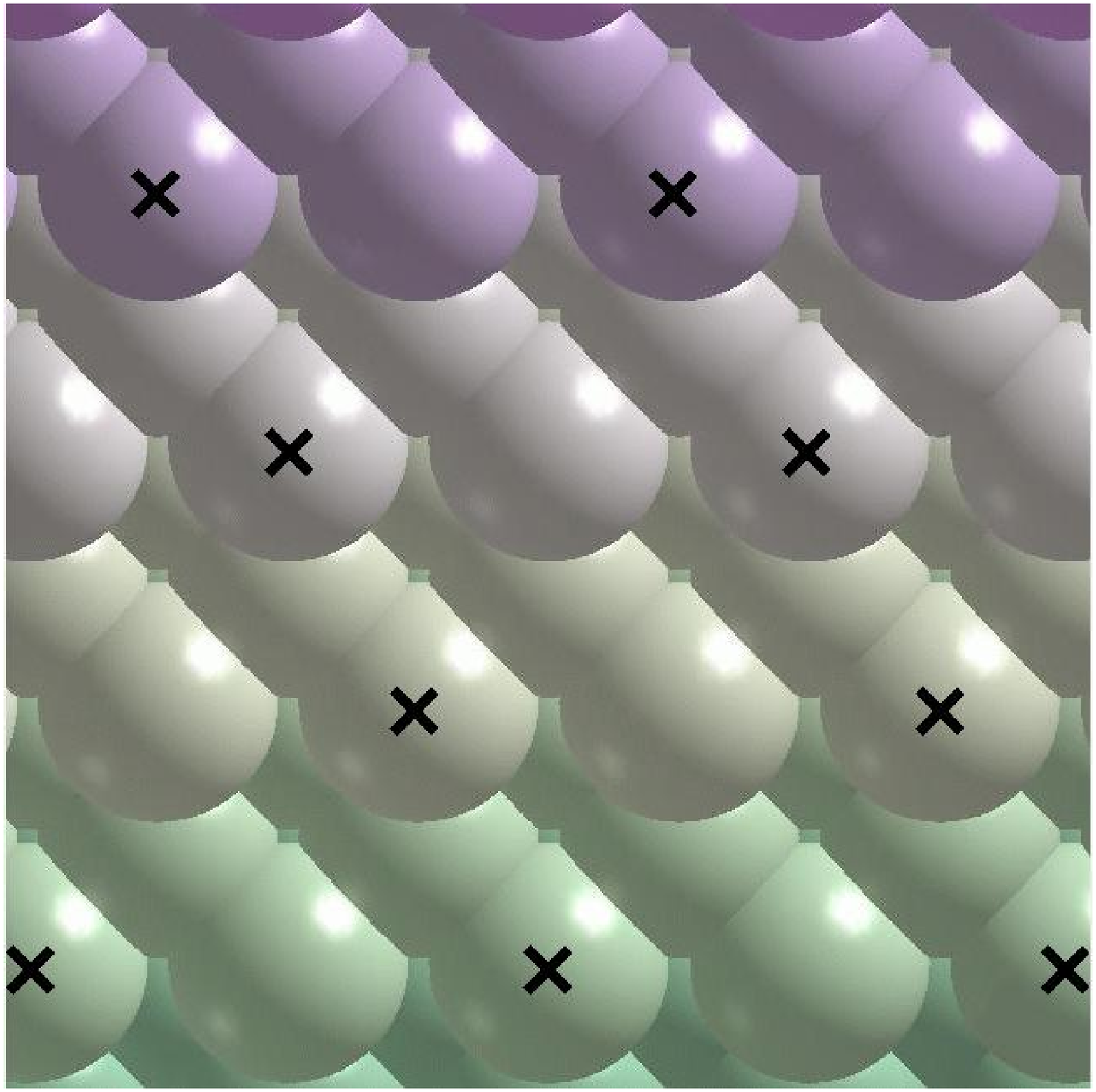}&
    \includegraphics[height=0.25\textwidth]{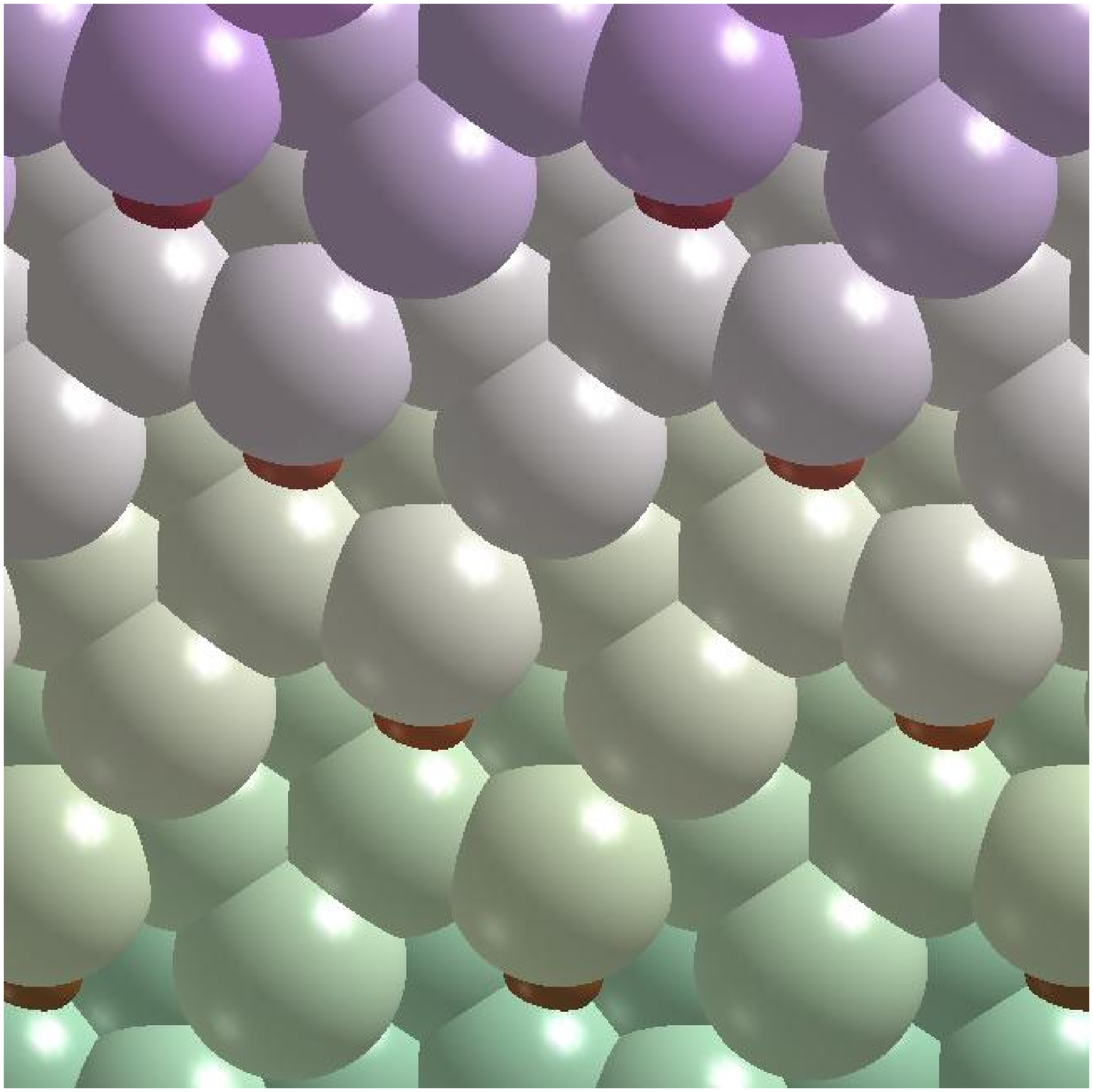}\\
    [0.5em]

    \multicolumn{2}{c}{\it O atom in tetrahedral-subsurface site}\\[-0.5em]
    \scriptsize\textsc{Initial structure} &
    \scriptsize\textsc{Optimised structure}\\
    \includegraphics[height=0.25\textwidth]{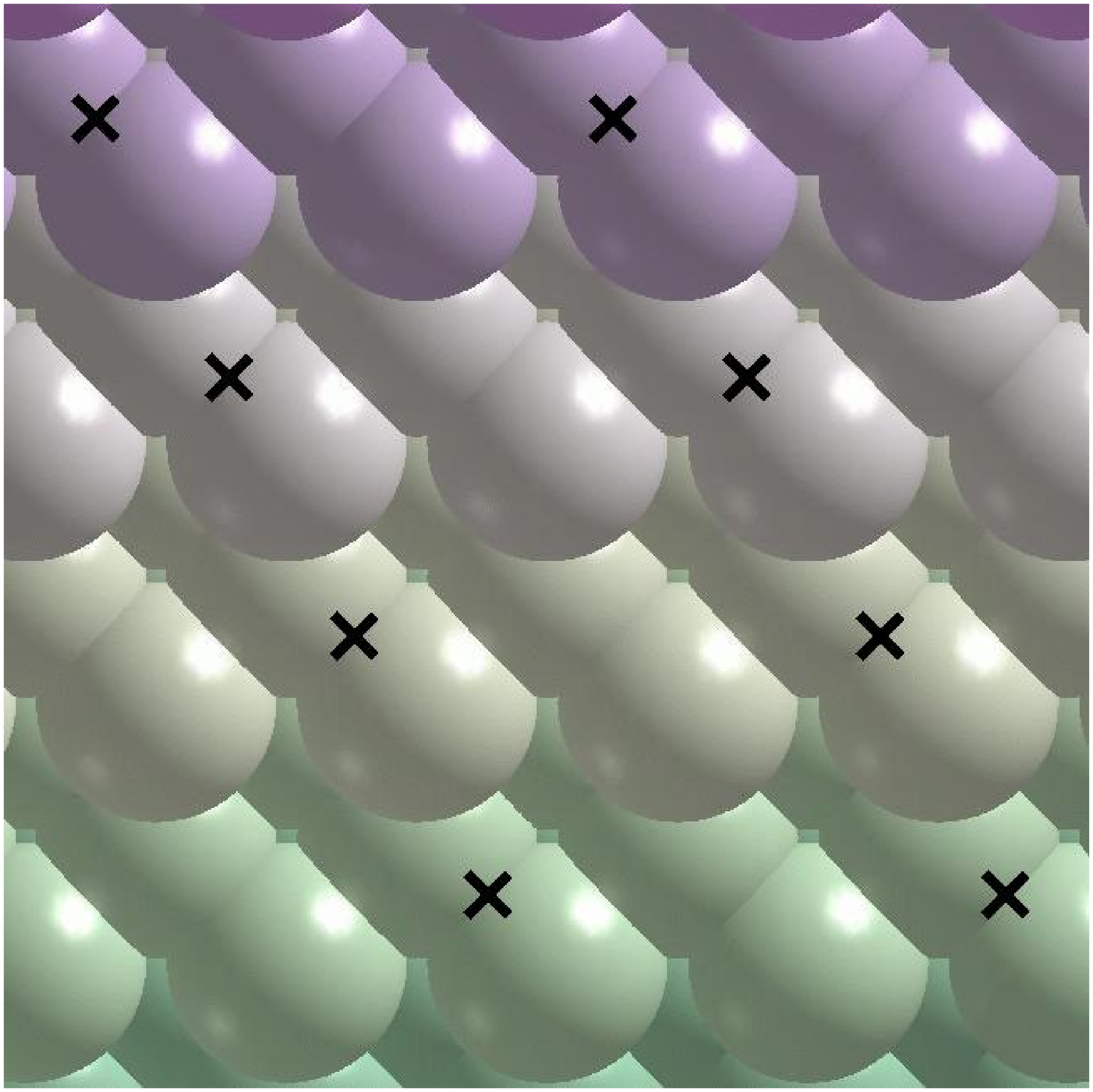}&
    \includegraphics[height=0.25\textwidth]{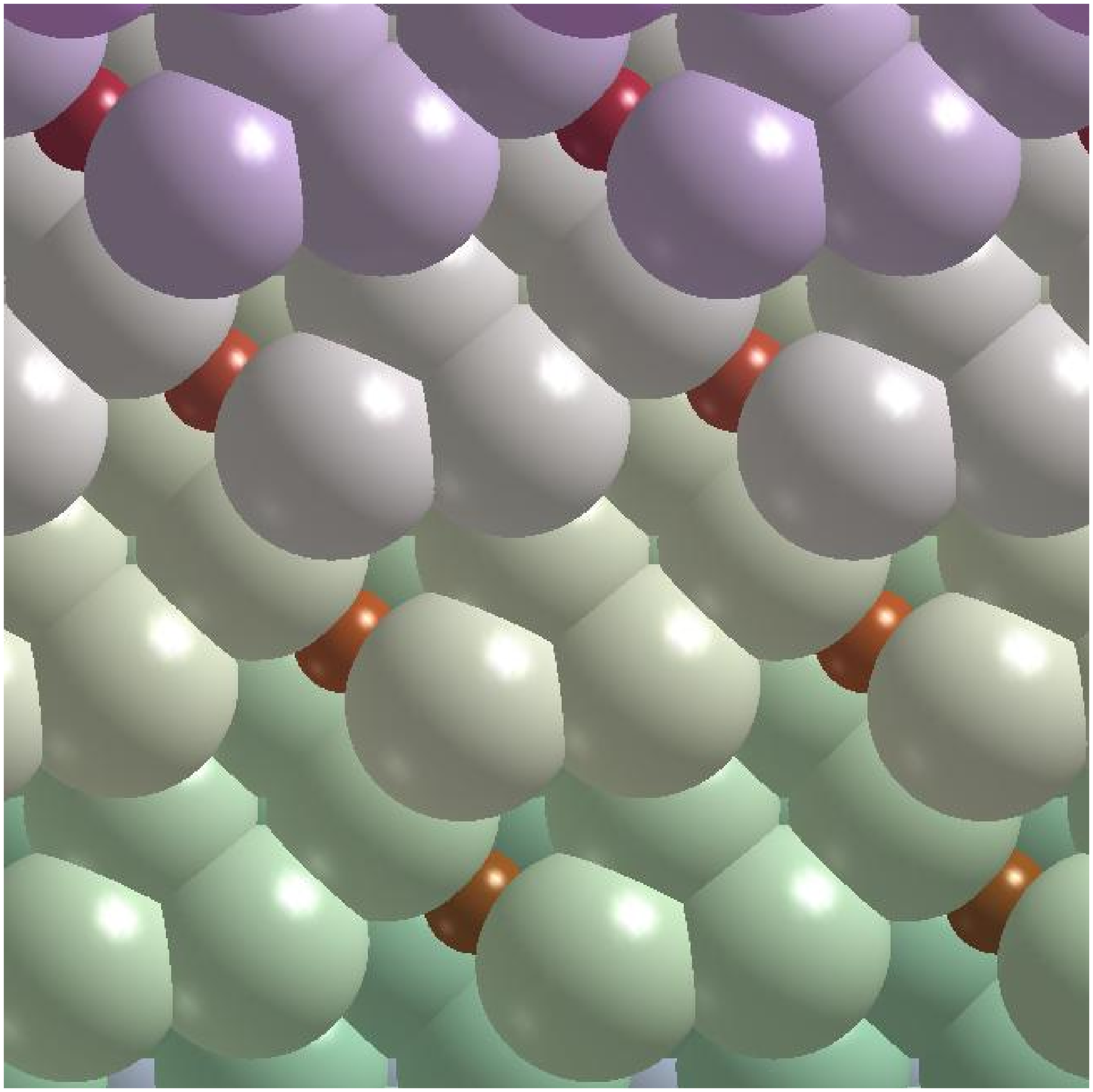}\\
  \end{tabular}
  \caption{Oxygen atoms located in subsurface sites on Ag(210)
    surface. Large (small) balls are silver (oxygen) atoms. On
    left-side panels the horizontal positions of subsurface-octahedral
    (\textsc{Octa}) (top left panel) and subsurface-tetrahedral
    (\textsc{Tetra}) (bottom left panel) sites are marked with crosses
    ($\mathbf{\times}$), while vertically these sites are located
    below the first layer Ag atoms in the octahedral and tetrahedral
    interstices. On right-side panels the corresponding optimised O/Ag
    structures are shown. Note a significant distortion of the
    substrate lattice---the (210) surface-structure attributes can
    hardly be recognised.}
  \label{fig:subsurface}
\end{figure}

\begin{figure}[htbp]
  \centering
  \linespread{1.0}
  \def\arraystretch{1.0}
  \begin{tabular}[t]{cc}
    \footnotesize \textsc{A--A + Octa}&
    \footnotesize \textsc{A--A + B}  \\[-0.3em]
    \includegraphics[height=0.25\textwidth]{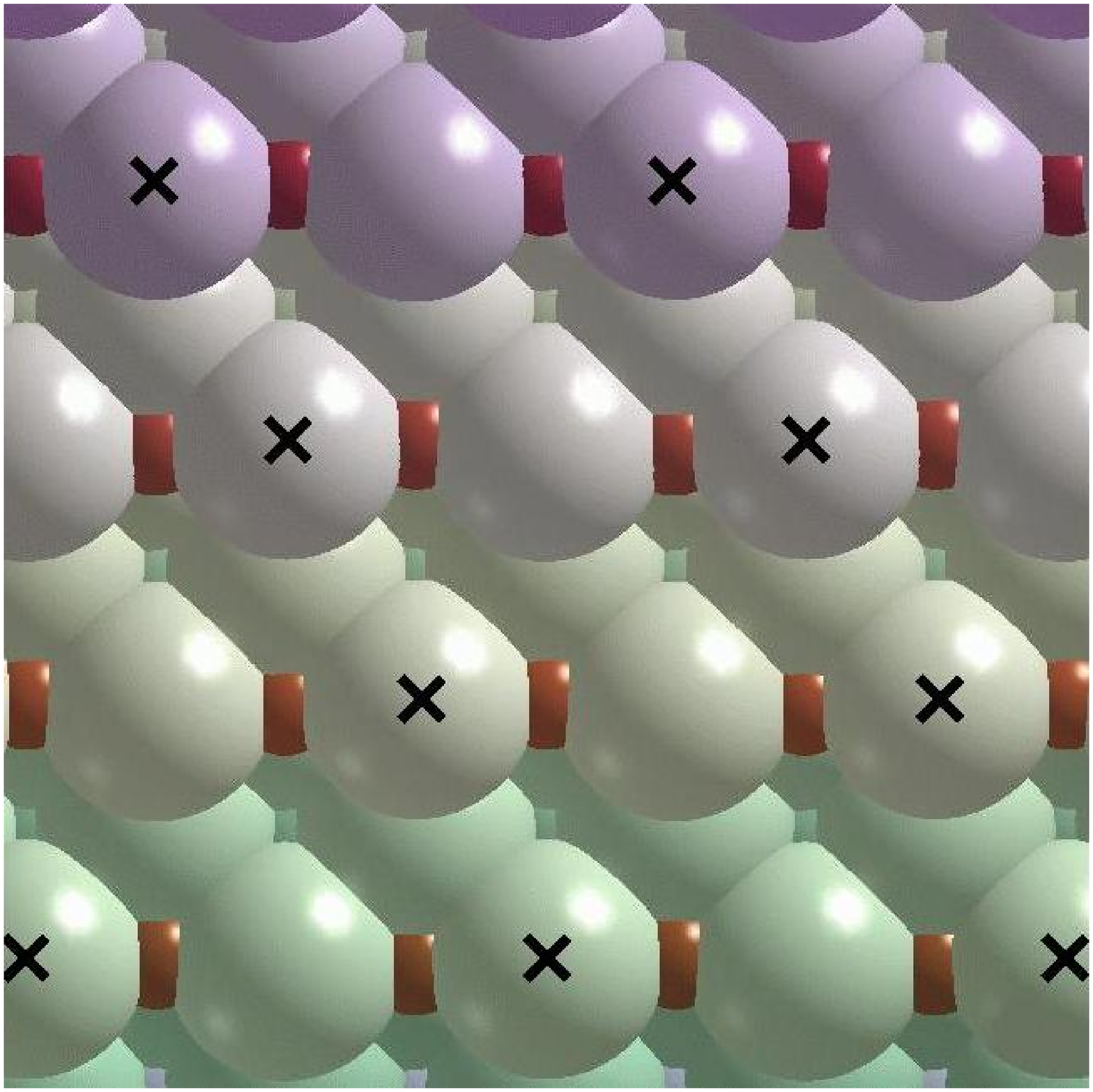}&
    \includegraphics[height=0.25\textwidth]{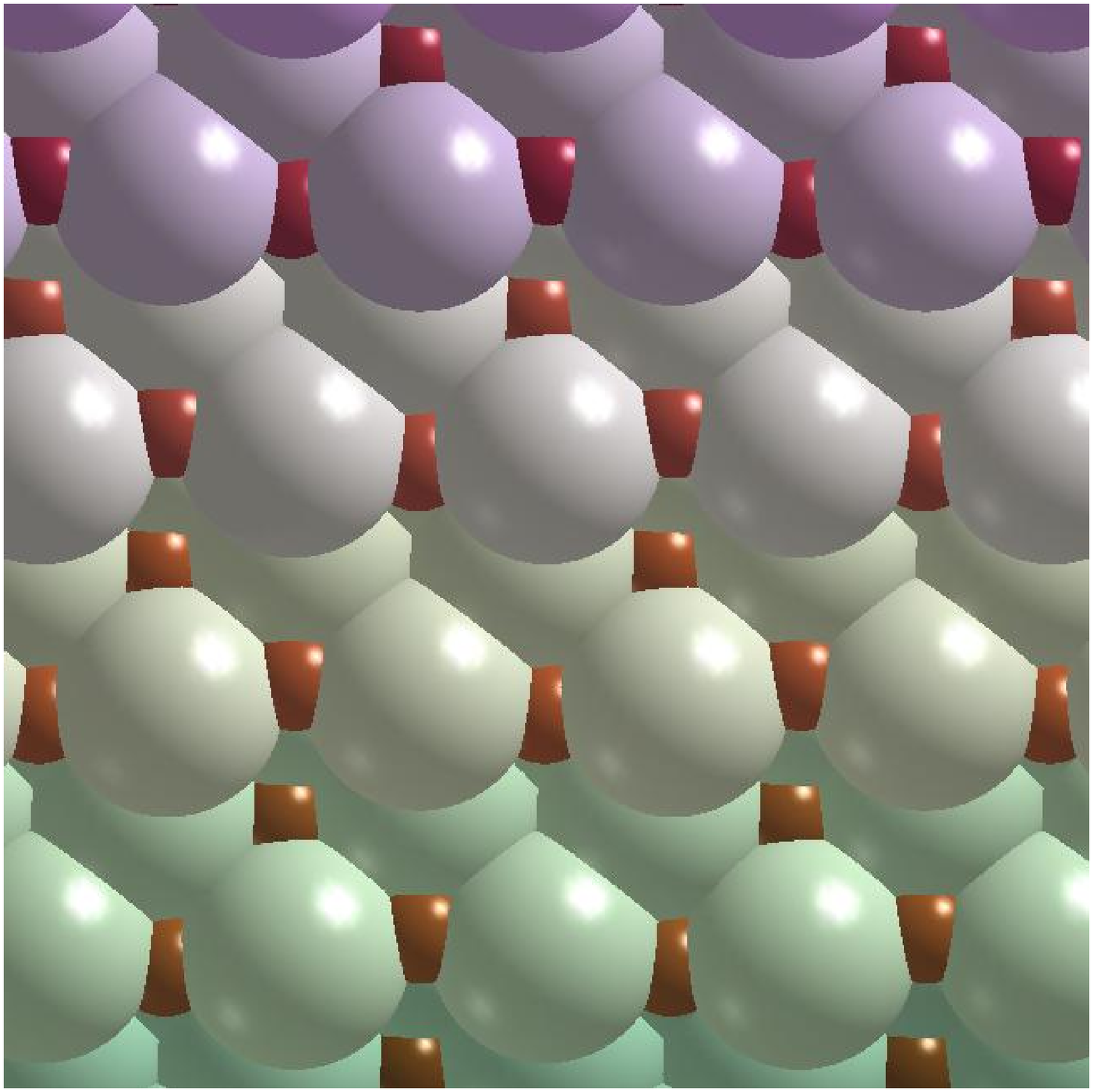}\\
  \end{tabular}
  \caption{The two competing O/Ag(210) structures at $\theta=3/4$~ML
    oxygen coverage. On both structures the oxygen atoms are
    decorating the step-edge (\textsc{A--A}). On the left structure
    an additional oxygen per ($2\times1$) super-cell is located in
    the subsurface-octahedral site (\textsc{Octa}), i.e., O atoms
    located straightly below the marked ($\mathbf{\times}$) step Ag
    atoms in octahedral interstices. On the right structure the
    additional O atom per ($2\times1$) super-cell is located in the
    hollow site below the step-edge (\textsc{B}). }
  \label{fig:mixed}
\end{figure}

\begin{figure}[htbp]
  \centering
  \linespread{1.0}
  \def\arraystretch{1.0}
  \begin{tabular}[t]{ccc}
    \footnotesize \textsc{A--A + Octa}&
    \footnotesize \textsc{A--A + B}&
    \footnotesize \textsc{A--A + T$_1$}  \\[-0.3em]
    \includegraphics[height=0.25\textwidth]{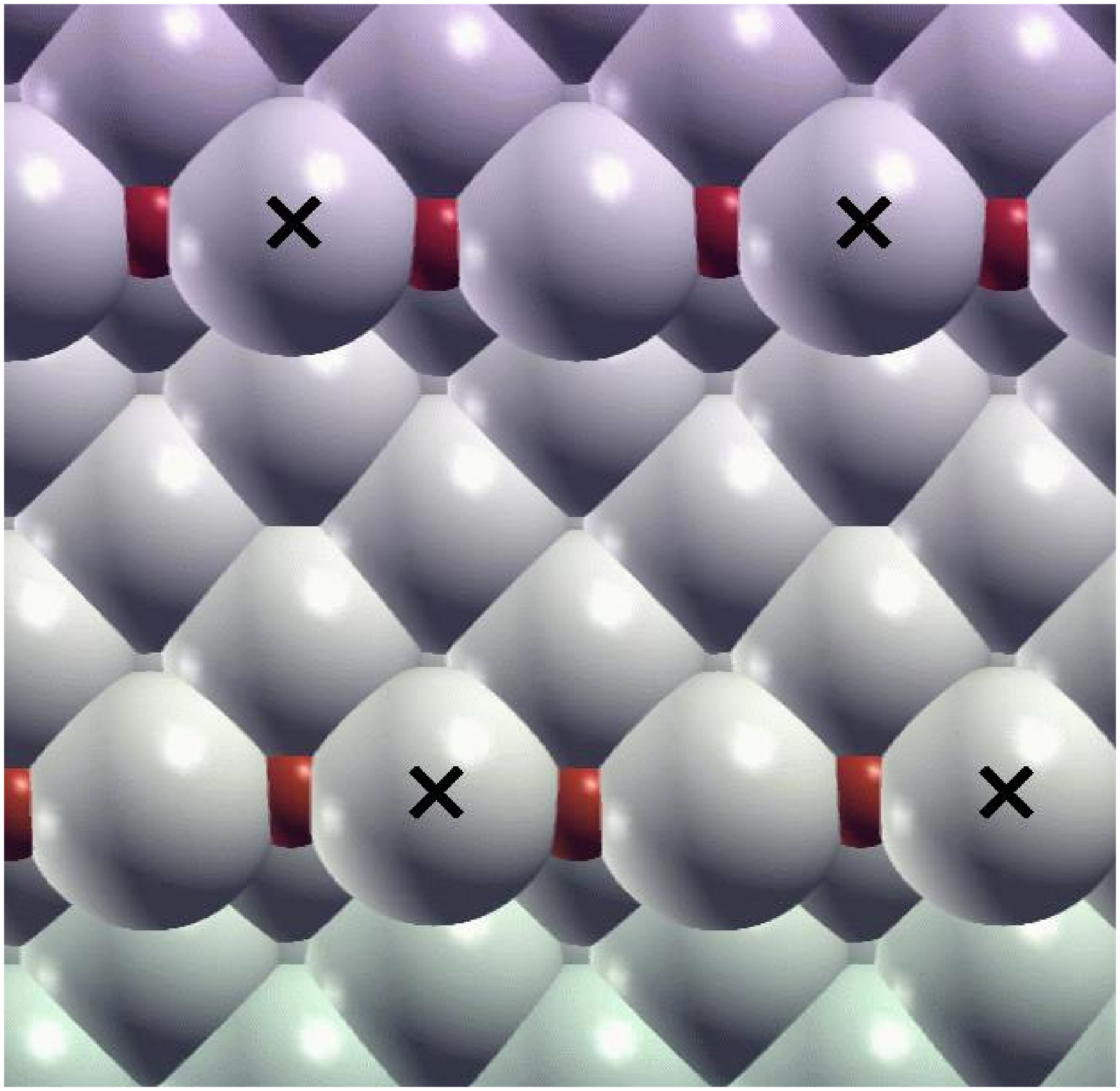}&
    \includegraphics[height=0.25\textwidth]{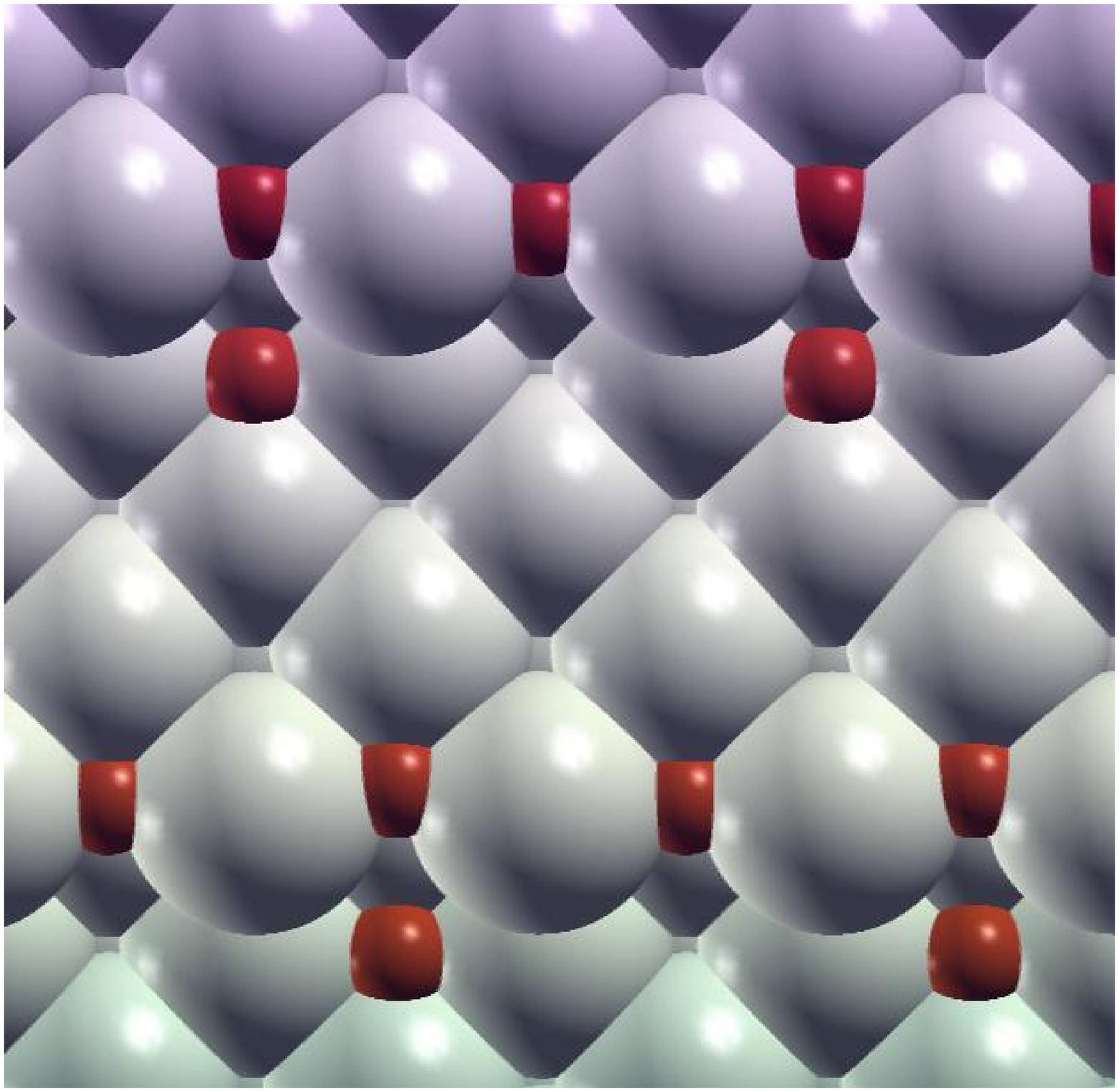}&
    \includegraphics[height=0.25\textwidth]{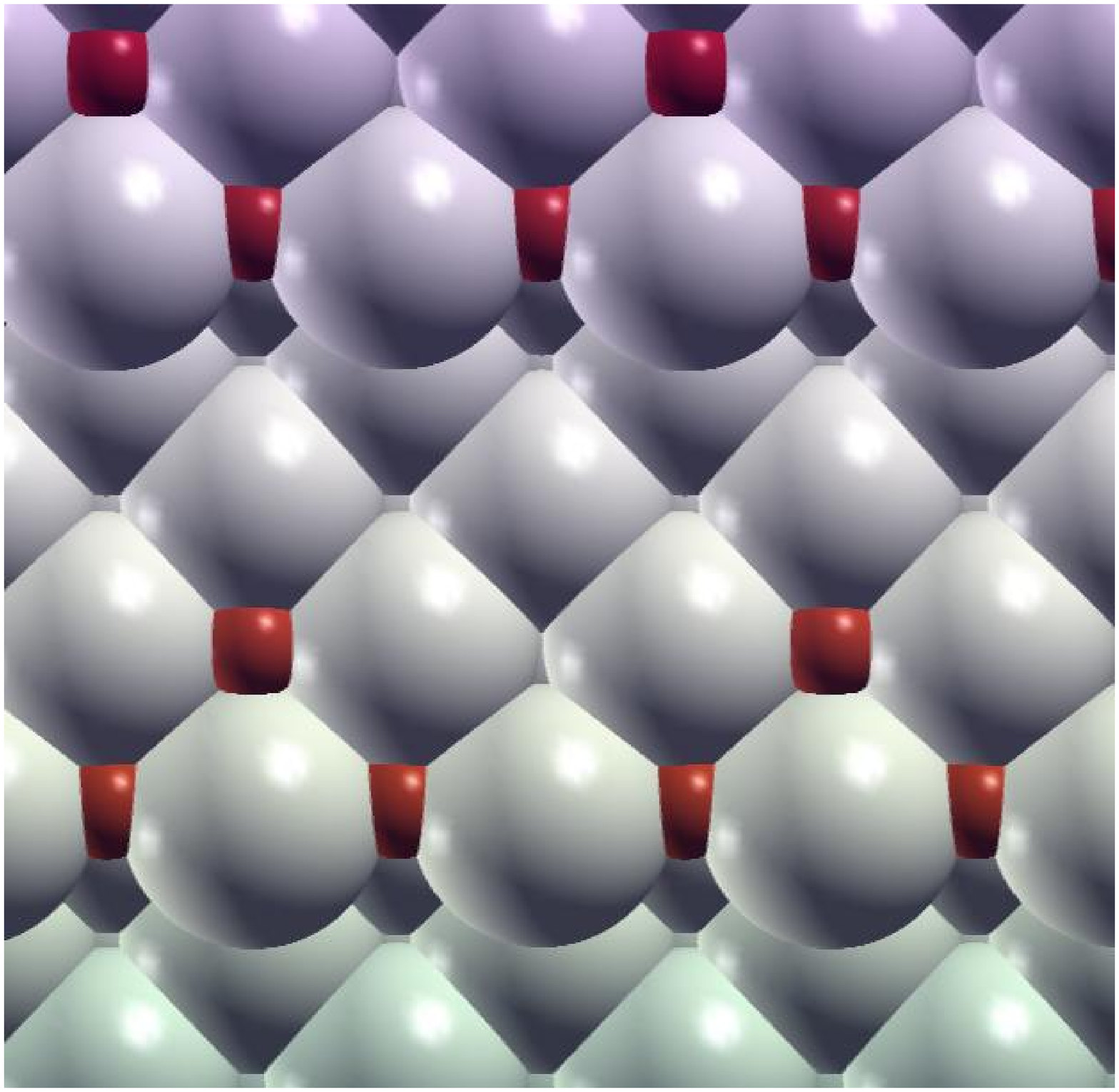}\\
  \end{tabular}
  \caption{The O/Ag(410) structures at $\theta=3/8$~ML oxygen
      coverage. From left to right: \textsc{A--A + Octa}, \textsc{A--A
        + B} and \textsc{A--A + T$_1$}. For the \textsc{A--A + Octa}
      structure the \textsc{Octa}-oxygen atoms are located straightly
      below the marked ($\mathbf{\times}$) step Ag atoms in octahedral
      interstices.}
  \label{fig:mixed410}
\end{figure}

The inclusion of oxygen into subsurface sites is strongly unfavourable
with respect to on-surface O adsorption, the resulting chemisorption
energies on Ag(210) surface being as low as $-0.07$~eV for
subsurface-octahedral site (\textsc{Octa}) and $-0.14$~eV for
subsurface-tetrahedral site (\textsc{Tetra}).\footnote{These
calculations were performed with Ag slabs composed of only 10
(210)-layers and somewhat smaller ($3\times5\times1$) Monkhorst-Pack
mesh.}  It is interesting to observe that the adsorption in
\textsc{Octa} and \textsc{Tetra} sites on Ag(100) at $\theta=0.25$~ML
\cite{O-Ag001} is considerably less stable (0.79 eV and 0.69 eV,
respectively) than on Ag(210) surface.  This indicates that the
presence of steps induces a stabilisation of subsurface oxygens.  In
the right panels of \Fig~\ref{fig:subsurface} we show top views of the
optimised structures with O atom in the \textsc{Octa} and
\textsc{Tetra} sites. Note that due to a significant distortion of the
substrate lattice induced by subsurface O atoms, the surface structure
of Ag(210) can hardly be recognised. However these subsurface sites
are significantly stabilised when a sufficient amount of O adatoms is
present on the surface.  The chemisorption energies for
\textsc{A--A+Octa} configuration\footnote{ In this configuration O
adatoms decorate the step-edge and an additional O atom per
($2\times1$) super-cell is located into \textsc{Octa} site situated
straightly below a step-edge Ag atom.}  are $-0.54$ and $-0.55$~eV for
Ag(210) and Ag(410) surfaces, respectively.  Also observe that the
substrate reconstruction is considerably reduced in these cases, and
that the Ag(210) and Ag(410) surface structures can be easily
recognised from the corresponding optimised O/Ag structures (see left
panel of Fig~\ref{fig:mixed} and right panel of
Fig~\ref{fig:mixed410}).

It is interesting to compare the stability of mixed on-surface +
subsurface configurations with the purely on-surface ones at a given
(high enough) oxygen coverage. In all the configurations we have
considered, O adatoms decorate the step-edge (A--A arrangement). An
additional oxygen atom is then considered to be located either into
subsurface \textsc{Octa} site (\textsc{A--A+Octa}) or adsorbed on a
given surface hollow site (see Figs~\ref{fig:mixed} and
\ref{fig:mixed410}). For Ag(210), we consider only the \textsc{A--A+B}
configuration with an additional O adatom located in the on-surface
hollow \textsc{B} site situated below the step edge. For Ag(410), in
addition to the \textsc{A--A+B} geometry, we also consider the
configuration with an additional O atom adsorbed on the terrace
\textsc{T$_1$} site (configuration \textsc{A--A+T$_1$}).  On Ag(210)
the mixed on-surface + subsurface \textsc{A--A+Octa} configuration is
more stable than the pure on-surface \textsc{A--A+B} configuration,
\echemAV\ being $-0.54$ and $-0.40$~eV, respectively.  On the
contrary, the opposite is true for the Ag(410) surface, where the
chemisorption energies for the \textsc{A--A+Octa} and \textsc{A--A+B}
configurations are $-0.55$ and $-0.68$~eV, respectively.  This
qualitative difference is related to the different O coverages of the
two surfaces, which is larger on Ag(210) surface ($\theta=3/4$~ML)
than on Ag(410) ($\theta=3/8$~ML).  This suggests that the
electrostatic repulsion between the adsorbates in the \textsc{A--A+B}
configuration is stronger on Ag(210) than on Ag(410) and makes this
geometry on Ag(210) significantly less favourable.  Also note that the
effect of high coverage in Ag(210) is weaker for the
\textsc{A--A+Octa} configuration essentially because the distance
between the adatoms is larger than in the \textsc{A--A+B} arrangement
and because the interaction involving subsurface O species is shielded
more effectively than for purely on-surface O atoms.  In addition to
this, we have found that on Ag(410) the incorporation of oxygen into
subsurface sites is less stable than on-surface adsorption also
because of the availability of stable terrace sites. The
\textsc{A--A+T$_1$} configuration is indeed more stable than the
\textsc{A--A+Octa} one.

%
%
\section{Conclusions}
\label{sec:Conclusions}
Our results are compatible with the suggestion, put forward in
Ref.~\cite{vattuone_PRL90}, that the HREELS peak found at 56 meV in
Ag(210), and missing in Ag(410), is due to atomic oxygen adsorbed
subsurface. Indeed our calculations, although performed at different
coverages, indicate that subsurface O is not stable in Ag(410), while
it is stabilised in Ag(210) at high enough coverage. Further
theoretical investigations spanning a larger range of coverages are
called for. Lattice vibrational calculations on selected O/Ag($n$10)
configurations would probably fully clarify the picture.

\section*{Acknowledgements}
This work has been supported by INFM ({\it Iniziativa trasversale
calcolo parallelo, Sezioni F e G}) and by the Italian MIUR (PRIN).
All numerical calculations were performed on IBM-SP3 and IBM-SP4
computers at CINECA in Bologna (Italy).

\bibliography{biblio}

\end{document}